\documentclass{bmcart}

\usepackage{amsmath}
\usepackage{braket}
\usepackage{graphicx}
\usepackage{slashed}
\usepackage{sidecap}
\usepackage{amsthm}
\usepackage{float}

\RequirePackage{natbib}
\usepackage{verbatim}
\usepackage{hyperref}
\usepackage[utf8]{inputenc} 
\usepackage{soul}

\startlocaldefs
\endlocaldefs

\begin{document}

\begin{frontmatter}

\begin{fmbox}
\dochead{Research}


\title{Italian Twitter semantic network during the Covid-19 epidemic}


\author[
   addressref={aff1,aff2},                   
  email={mattia.mattei14@libero.it}
 ]{\inits{MM}\fnm{Mattia} \snm{Mattei}}
\author[
   addressref={aff3},
]{\inits{GC}\fnm{Guido} \snm{Caldarelli}}
\author[
   addressref={aff2},
]{\inits{TS}\fnm{Tiziano} \snm{Squartini}}
\author[
   addressref={aff2, aff4},
  ]{\inits{FS}\fnm{Fabio} \snm{Saracco}}


\address[id=aff1]{
  \orgname{University of Salento}, 
  \street{P.zza Tancredi 7},                     %
  \postcode{73100}                                
  \city{Lecce},                              
  \cny{Italy}                                    
}
\address[id=aff3]{%
  \orgname{``Ca' Foscari'' University of Venice},
  \street{Dorsoduro 3246},
  \postcode{30123}
  \city{Venice},
  \cny{Italy}
}
\address[id=aff2]{%
  \orgname{IMT School for Advanced Studies},
  \street{P.zza S. Ponziano 6},
  \postcode{55100}
  \city{Lucca},
  \cny{Italy}
}
\address[id=aff4]{%
  \orgname{Institute for Applied Computing ``Mauro Picone'' (IAC), National Research Council},
  \street{via dei Taurini 19},
  \postcode{00185}
  \city{Rome},
  \cny{Italy}
}


\begin{artnotes}
\end{artnotes}

\end{fmbox}


\begin{abstractbox}
\begin{abstract}
The Covid-19 pandemic has had a deep impact on the lives of the entire world population, inducing a participated societal debate. As in other contexts, the debate has been the subject of several d/misinformation campaigns; in a quite unprecedented fashion, however, the presence of false information has seriously put at risk the public health. In this sense, detecting the presence of malicious narratives and identifying the kinds of users that are more prone to spread them represent the first step to limit the persistence of the former ones. In the present paper we analyse the semantic network observed on Twitter during the first Italian lockdown (induced by the hashtags contained in approximately 1.5 millions tweets published between the 23rd of March 2020 and the 23rd of April 2020) and study the extent to which various discursive communities are exposed to d/misinformation arguments. As observed in other studies, the recovered discursive communities largely overlap with traditional political parties, even if the debated topics concern different facets of the management of the pandemic. Although the themes directly related to d/misinformation are a minority of those discussed within our semantic networks, their popularity is unevenly distributed among the various discursive communities.


\end{abstract}




\end{abstractbox}
%

\end{frontmatter}





\section{Introduction}

The Covid-19 pandemic has had a deep impact on nearly every human activity; as such, it is not surprising that it has generated a widespread online debate which, in turn, attracted the interest of scholars from different disciplines~\cite{Rovetta2020,Celestini2020, Gallotti2020, Cinelli2020,yang2020covid19, Caldarelli2020b,Patuelli2021}. Due to its unmediated nature, the online debate was affected by quite an amount of low-quality contents that, more than in other circumstances, had the potential to severely put at risk the public health. In this sense, the High Representative of the Union for Foreign Affairs and Security Policy explicitly exposed her concerns about the possible effects of a wrong communication, on social media, during the pandemic: ``Disinformation can have severe consequences: it can lead people to ignore official health advice and engage in risky behaviour, or have a negative impact on our democratic institutions, societies, as well as on our economic and financial situation.''~\cite{HRUFA2021}.

Indeed, the already prolific research on the diffusion of d/misinformation~\cite{Gonzalez-Bailon2013,cresci2015fame, Stella2018,Ciampaglia2018,Ferrara2016rise,Ferrara2019,Cresci2019WebSci,Bovet2019, Becatti2019, Caldarelli2020a} was enriched by the reactive attention of scholars. For instance, Gallotti et al.~\cite{Gallotti2020} released a real-time dashboard to monitor the risk of exposure to d/misinformation in the various countries on Twitter and proposed an Infodemic Risk Index, based on epidemic studies. In particular, they were able to detect early-warning signals related to the diffusion of d/misinformation campaigns. Interestingly, Celestini et al.~\cite{Celestini2020} analysed the Italian debate on Facebook and found that controversial topics associated to known false information represented a limited amount of the traffic and, moreover, are less able to engage the audience than reliable media. Even more remarkably, authors observed the presence of a sort of small-world effect in the exposure to URLs, such that nearly any user can access any kind of news source, even if the navigation is limited to a reduced number of pages. Following a similar line of research, Yang et al.~\cite{yang2020covid19} analysed both Twitter and Facebook, focusing their attention on links to low-credibility contents; in particular, they observed the presence of a limited number of extremely influential accounts, i.e. the d/misinformation \emph{superspreaders}. 

The comparison of several platforms is also the research target of~\cite{Cinelli2020}: Gab, Facebook, Instagram, Reddit and Youtube were analysed and, by fitting the information spreading via an epidemic model, an $R_0$ parameter was assigned to any platform. Even if the spreading patterns are similar, the various online social networks are differently exposed to the risk of d/misinformation.

Across the entire literature overview, a limited attention is paid to identify the narratives shared by the various users. In this sense, the contribution of \cite{Radicioni2020,radicioni2021networked} is twofold: first, the authors attempt at inferring the various communication strategies on Twitter by leveraging on the different usage of the hashtags made by the users (and considering both the attitude of the various accounts towards the use of different hashtags and the popularity of the latter ones); second, they represent one of the few examples of applications of the entropy-based framework~\cite{Squartinia,Cimini2018} for the analysis of Online Social Networks. So far, in fact, entropy-based null-models have been successfully employed to analyse economic~\cite{squartini2011analytical, Mastrandrea2014, Saracco2015a, Saracco2016a, Pugliese2019} and financial networks~\cite{Squartini2013, Gualdi2016a}, either to reconstruct such systems from limited information~\cite{SQUARTINI20181} or as a benchmark for the analysis of their network structure. Only recently, these methods have been applied to the analysis of Online Social Networks, evidencing non-trivial phenomena such as customer tastes in online retail networks~\cite{Becatti2018}, the presence of discursive communities on Twitter~\cite{Becatti2019}, the presence of coordinated activities of automated accounts~\cite{Caldarelli2020a}, the extent to which the various discursive communities are exposed to d/misinformation campaigns~\cite{Caldarelli2020b}.

In the present manuscript, we examine the Italian online debate during the peak of the first wave of the Covid-19 pandemic, following the approach of~\cite{Radicioni2020,radicioni2021networked}, i.e. focusing our attention on the semantic side. To this aim, we represent the Twitter online debate as a bipartite network in which the two layers respectively represent accounts and hashtags and a link connect two nodes if the considered account has used at least once the given hashtag during the observation period. Then, we extract the semantic network of hashtags by employing an entropy-based null-model as a benchmark to project the bipartite system into a monopartite one~\cite{Saracco2016}. Differently from~\cite{Radicioni2020,radicioni2021networked}, where the focus was on the communication strategies of the different discursive communities (respectively, during the 2018 Italian electoral campaign and about the debate on migration policies), here we produce a single semantic network across the entire period of the debate. The various communities engage in different discussions in different moments, depending on the interest towards the topic. Nevertheless, as also observed by focusing on URLs news sources~\cite{Caldarelli2020b}, the various discursive communities are differently exposed to dis/mis-information campaigns.

Following a previous study on a similar data set~\cite{Caldarelli2020b}, the discursive communities largely overlap with political parties. As already commented in~\cite{Caldarelli2020b}, this behaviour is probably due to the pre-existence of discursive communities that shape and condition the platform environment. Otherwise stated, the debate about the spread of coronavirus and the efficiency of the adopted countermeasures developed in a context in which various discursive communities are already present: if a users is following a group of accounts, those will be the group from which she/he will receive updates and with which she/he will engage in discussions. In the present paper, as well as in previous ones by the same authors, we will avoid the term \emph{echo chamber}, due to its different usage in the literature: while in~\cite{Cinelli2021}, Twitter echo chambers are defined via the patterns of news consumption (news outlets are given a political orientation and the user is given a membership based on the her/his news consumption), here we follow the approach in~\cite{Becatti2019,Caldarelli2020a,Caldarelli2020b, Radicioni2020} and avoid any \emph{a priori} labeling; instead, we focus on the interactions among users whence our choice of calling them \emph{discursive communities}.

The presence of politically-oriented discursive communities is reflected in the structure of the discussion, mimicked by the structure of our semantic network. Remarkably, the hashtags referring to d/misinformation arguments represents a minority of the entire semantic network, confirming the observations in~\cite{Celestini2020}. Interestingly, not all users are exposed in the same way to those hashtags: as already observed in~\cite{Caldarelli2020b}, right-wing discursive communities use more hashtags related to d/misinformation.\\

The manuscript is organised as follows: we present the main results in Section \ref{sec:res}, discuss the various discursive communities in Subsection~\ref{ssec:DiscComm} and analyse the semantic network in Subsection~\ref{ssec:SemNet}. Then, we briefly present the methodology used in Section~\ref{sec:Meth} and conclude by discussing our findings in Section~\ref{sec:Disc}.

\section{Results}\label{sec:res}

\subsection{Identification of the discursive communities}\label{ssec:DiscComm}

Users can interact on Twitter in different ways: for example, one can \emph{retweet} the content of another user, hence endorsing it~\cite{Conover2011} and raising the content visibility; in order to infer the membership of the various accounts, in the present paper we leverage on this activity, following the procedure adopted in~\cite{Becatti2019,Caldarelli2020a}.

\subsubsection{Discursive communities of verified users}

On Twitter there are essentially two kind of accounts: the ones that are verified and whose authenticity is certified by Twitter itself - and belonging to journalists, politicians, VIPs or being the official accounts of ministries, political parties, newspapers and TV-channels - and the ones that are not verified. About the former ones, we have the largest available information: interestingly enough, verified accounts are more devoted to product original posts than sharing existing ones~\cite{Caldarelli2020b}. Indeed, these accounts act like seeds, proposing new arguments for the public debate.

First, we divided the users of our data set into two groups: the verified and the non-verified accounts. Then, we represented the system as a bipartite network, where verified users are gathered on one layer and the non-verified users are gathered on the other; an edge between vertices of different layers indicates that one has retweeted the other's content at least once during the period of study. To infer the membership of the verified users to a certain discursive community, we projected the bipartite network onto the layer of verified accounts. The procedure consists in counting, for each pair of verified accounts, how many non-verified users have retweeted both of them. 
The rationale is the following: the largest the number of non-verified users interacting (via tweet or retweet) with the same couple of verified accounts, the greater the possibility that the two are perceived as similar by the audience of unverified ones~\cite{Becatti2019}. Nevertheless, the sole information regarding the number of common neighbors is not enough to state if the two verified accounts are similar: two users may have a great number of common neighbours just because they are both extremely active on Twitter or because their nearest neighbours are among the most active unverified accounts. The statistical significance of the number of common neighbours between two verified accounts can be evaluated by comparing it with its expected value, according to a null-model~\cite{Saracco2016}; once the amount of common nearest neighbours is deemed as statistically significant, we can connect the considered couple of nodes in the projected network.

In the present case, the adopted benchmark is the entropy-based null-model constraining the degree sequences of the bipartite network i.e. the Bipartite Configuration Model (\emph{BiCM}~\cite{Saracco2015a}). The details about the whole procedure, i.e. the null-model construction and the validation, can be found in Section~\ref{sec:Meth}.\\

The result of the projection is a monopartite network of 3,786 edges and 576 different verified users; we used the Louvain algorithm~\cite{Blondel2008} to detect the various communities\footnote{Respect to the standard Louvain algorithm, using the definition of the Modularity using the Chung-Lu null-model~\cite{Chung2002}, in the present paper we used the monopartite entropy-based configuration model~\cite{squartini2011analytical}. Literally, Chung-Lu null-model can be considered as a sparse matrix approximation of the null-model in~\cite{squartini2011analytical} and, indeed, returns wrong results in the presence of strong hubs~\cite{Cimini2018}.}. The algorithm provided different communities of verified users with an overall modularity equal to 0.61. Discursive communities with a clear political orientation were already detected in other works~\cite{Becatti2019,Caldarelli2020a} but the arguments studied there (i.e. the 2018 Italian elections and the political debate about migration policies) were political in nature. Remarkably, as already observed in~\cite{Caldarelli2020b}, the political discursive communities shape even the wider debate targeting the Covid-19 pandemic - and including different health, scientific, societal, economic and political facets. In order to gain more insight on the partition and spot the presence of sub-communities, we re-run the Louvain algorithm inside each one of the largest groups: as a result, we individuated five major modules that can be associated with the main Italian political parties (more details on the Italian political scenario and the identity of the verified users mentioned below can be found in Appendix \ref{app:polcomm}):\\

\begin{itemize}
    \item The \textbf{M5S} community contains 85 accounts of supporters and politicians of the \emph{Movimento 5 Stelle} party. This community includes the official account of the movement and the accounts of personalities like \emph{Beppe Grillo}, \emph{Luigi Di Maio} and \emph{Virginia Raggi}. Interestingly the account of the former premier \emph{Giuseppe Conte} is in this community. There are also some official accounts of ministries (like \emph{Ministero della Giustizia} and \emph{Ministero del Lavoro}\footnote{Ministry of Justice and Ministry of Labour.}) and newspapers and TV-channels like \emph{Il Fatto Quotidiano} and \emph{Report Rai 3}.
    \item The right-wing (\textbf{DX}) community is constituted by the supporters and the politicians of the right-wing parties \emph{Lega Nord} and \emph{Fratelli d'Italia}. This is much smaller than the previous one (only 32 elements) and contains the accounts of \emph{Matteo Salvini}, \emph{Giorgia Meloni}, \emph{Lorenzo Fontana}, \emph{Vittorio Sgarbi} and the \emph{Russian embassy}.
    \item The Democratic Party (\emph{Partito Democratico}, or \textbf{PD}) community contains politicians and supporters of the main center-left party. It contains 37 nodes among which the official accounts of politicians as \emph{Nicola Zingaretti}, \emph{Paolo Gentiloni}, \emph{Enrico Letta} and the party official one (\emph{pdnetwork}).
    \item The Italia Viva (\textbf{IV}) group contains accounts of politicians affiliated to the homonym center-left wing party. Here we can find the official account of the party and the accounts of \emph{Matteo Renzi}, \emph{Ivan Scalfarotto} and \emph{Maria Elena Boschi}. Interestingly, we also signal the presence of \emph{Roberto Burioni}, one of the most popular Italian virologists nowadays, particularly active on the popularization on subjects related to the pandemic. The group contains 24 accounts in total.
    \item The Forza Italia (\textbf{FI}) community contains only 11 accounts and all of them are of politicians affiliated to the \emph{Forza Italia} center-right wing party; for example, it contains accounts like \emph{Silvio Berlusconi}, \emph{Antonio Tajani} and \emph{Renato Brunetta}.\\
\end{itemize}

\begin{figure}[t!]
    \centering
    \includegraphics[scale=0.6]{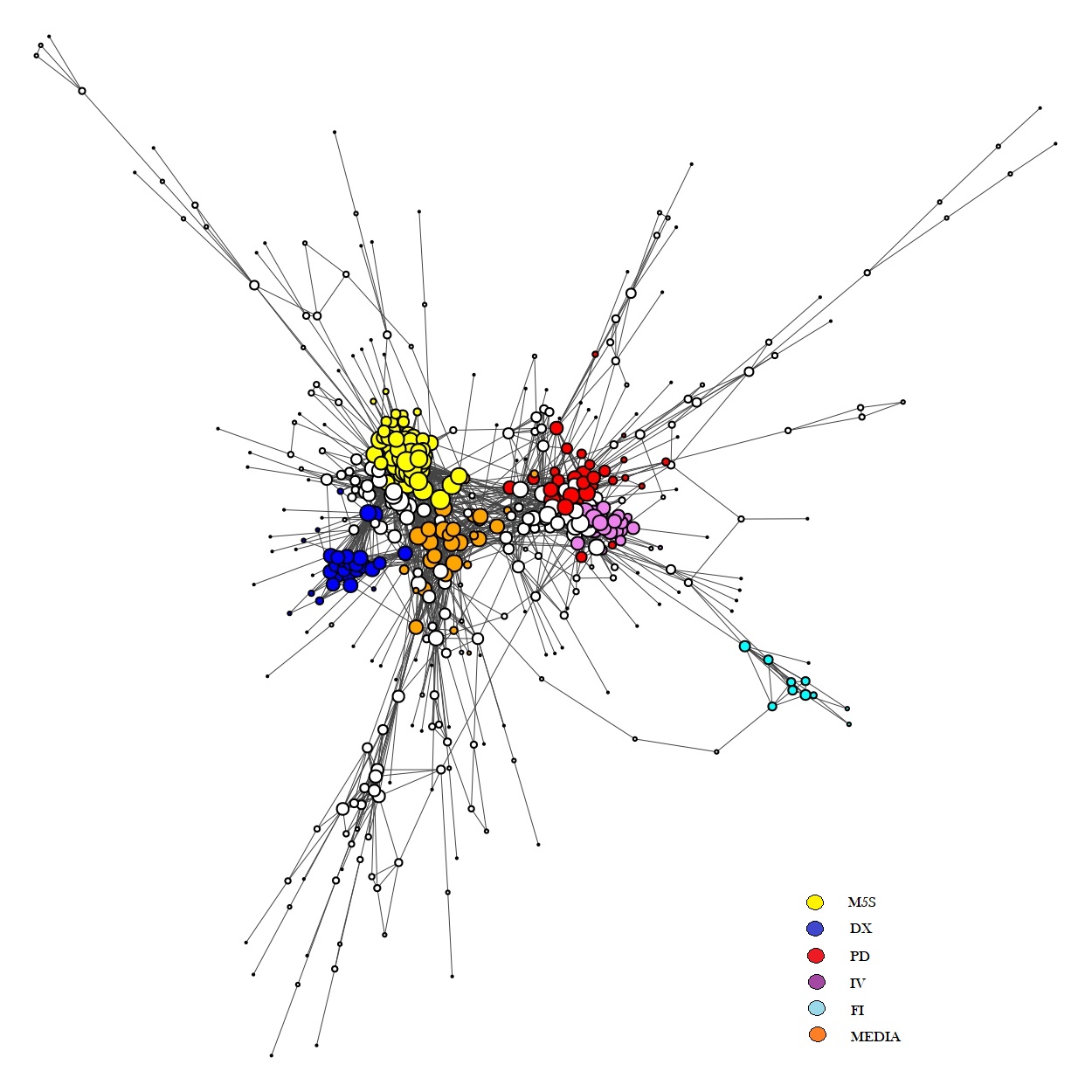}
    \caption{\textbf{The Largest Connected Component (LCC) of the validated network of verified accounts} Two verified users are connected if they share a statistically significant number of neighbours in the bipartite network, i.e. if a sufficiently high number of non-verified users has retweeted both of them compared with the expections of the BiCM~\cite{Saracco2015a}. The main communities described in the text are pictured with different colors, nodes belonging to smaller communities are in white. The network is represented using Fruchterman-Reingold layout.}
    \label{Figure1}
\end{figure}

In addition to the political groups described above, we also considered the \textbf{MEDIA} community, i.e. the community that contains the official accounts of newspapers (like \emph{Repubblica} and \emph{Agenzia Ansa}), TV-channels (like \emph{La7TV}), radio and other media. This group contains 33 verified accounts. The Largest Connected Component (LCC) of the validated network of verified users is shown in Fig.~\ref{Figure1}. The main communities are depicted with different colors.

\subsubsection{Political orientation of non-verified users}

Once verified accounts are associated to the various discursive communities, following the approach of.~\cite{Conover2011}, we can infer the membership of unverified  ones by considering their interactions in the retweet network. As in~\cite{Caldarelli2020a,Radicioni2020,Caldarelli2020b}, we use the membership of verified users as (fixed) seeds for the label propagation proposed by Raghavan et al.~\cite{Raghavan2007b}. Let us remind that in case this algorithm cannot find a dominant label for a specific vertex (in case of a tie), it randomly removes some of the edges attached to that vertex and repeats the procedure; due to its intrinsic stochasticity, we run the label propagation 500 times and assigned to each node the most frequent label (actually, the noise in the assignment of the labels is extremely limited): as a result, approximately 89\% of the users in the network have been inserted in one of the 6 discursive communities described in the subsection above. They are distributed as follows:

\begin{itemize}
\item 117,798 users in MEDIA group;
\item 27,989 users in  DX group;
\item 7,230 users in  M5S group;
\item 1,685 users in  IV group;
\item 1,408 users in  PD group;
\item 430 users in FI group.
\end{itemize}

As expected, the MEDIA community is the biggest one: it represents users who considerably share news from the accounts of newspapers, radios or newscasts. Looking at the political groups, it is interesting to see that the M5S group contains less elements than the DX community: by considering only verified accounts, the M5S community includes more than 1.5 times the total number of users of the DX one. The center-left wing (PD and IV) and FI communities are quite small; the vertices with the largest degree are mostly verified accounts, belonging to the MEDIA group (e.g. newspapers or press agencies as \emph{La Repubblica}, \emph{La Stampa}, \emph{Ansa}).

The most retweeted accounts are those of \emph{Giorgia Meloni} (DX community) and \emph{Roberto Burioni} (IV community). Remarkably, the vertex with the largest degree, that the label propagation algorithm assigns to DX community, is a non-verified user whose number of neighbors amounts at 27,000 and whose activity is that of sharing news everyday\footnote{For privacy reasons, we are allowed to mention the screen name of verified accounts only.}. It often shares racially-motivated news that the debunking web-site \href{https://www.bufale.net/}{Bufale.net} has identified as lacking in sources.

Overall, our results confirm the ones observed in other works, i.e. that communication on Online Social Networks (OSNs) is characterized by a strong polarization, in turn inducing a strongly modular system~\cite{Adamic2005,Conover2011,DelVicario2016c,Zollo2017,Bovet2019,Becatti2019,Caldarelli2020a,Caldarelli2020b, Radicioni2020}.

\subsubsection{Social-bots}

Social bots, or simply bots, are social accounts governed - completely or partly - by pieces of software that automatically create, share and like contents on Twitter and other platforms. In general, the usage of automated accounts is allowed by Twitter platform for promotional purposes by various companies  (see Twitter Developer's \href{https://help.twitter.com/en/rules-and-policies/twitter-automation}{Automation Rules}). Nevertheless, bots often pretend to be human accounts and aim at influencing and diverting the course of discussions by inflating the visibility of some genuine accounts~\cite{cresci2015fame,Ferrara2019}.

In the present manuscript we used \emph{Botometer}~\cite{Ferrara2019}, a tool based on supervised machine learning: given a Twitter account, Botometer extracts over 1,000 features and produces a classification score called ``bot score'': according to the algorithm, the higher the score, the greater the likelihood that the account is controlled completely or in part by a software. Botomoter revealed that in our data set social-bots shared 52,054 different tweets (approximately the 3\%-4\% of the entire data set) with 74,884 hashtags. The most used hashtags by bots are: \#\emph{iorestoacasa}, \#\emph{italia}, \#\emph{news}, \#\emph{quarantena} and \#\emph{conte}\footnote{`stay at home', `italy', `news', `quarantine' and Giuseppe `Conte', the former Prime Minister.}.

\begin{figure}
    \centering
    \includegraphics[scale=0.8]{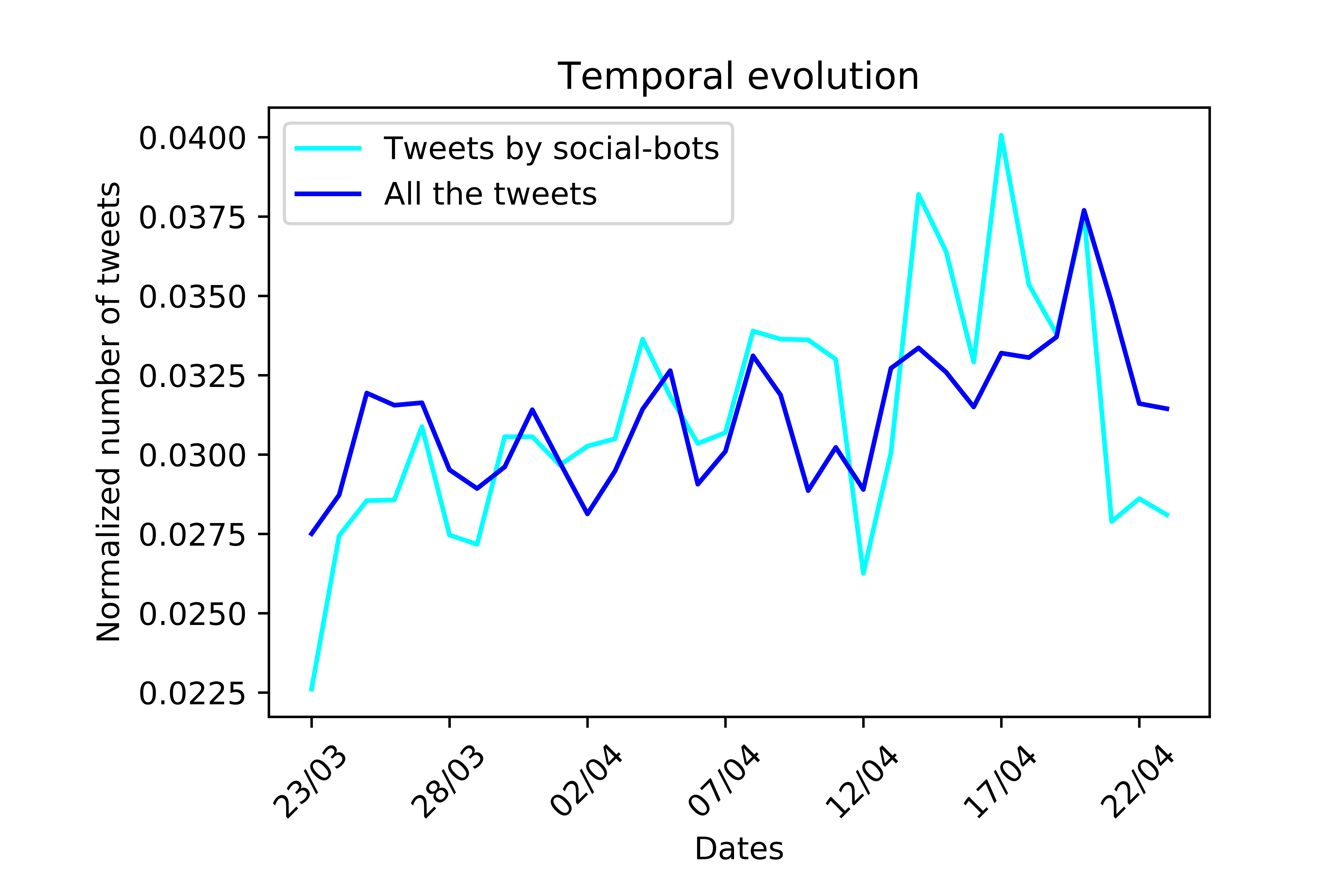}
    \caption{
      \textbf{The temporal evolution of the number of tweets shared by social-bots (cyan line), as compared to the analogous trend for all users (blue line)} The values are normalised by the total number of tweets sent in the entire period by the considered set of accounts. Remarkably, the automated accounts' trend shows two peaks on the 14th and on the 17th of April, which are not present in the analogous trend for the entire set of users in our data.}
    \label{Figure3}
\end{figure}

In Fig.~\ref{Figure3} the temporal evolution of the number of tweets shared by social-bots during the period of study is displayed. Remarkably, this trend is similar to the trend for the entire set of users, except for the presence of two peaks, on the 14th and on the 17th of April. In this period, the political debate in Italy became more intense, in particular about the European Stability Mechanism (ESM). The ESM is an international organization born as a European financial fund for the financial stability of the euro area; in those days the Italian government was considering the possibility of using these funds in order to limit the impact of the pandemic, analysing the possible consequences. In particular, right-, center-right-wing parties and M5S were against its usage, while both PD and IV were in favour of it. Then, we looked at the interactions of social bots with the political groups identified before; in particular, we saw how many bots retweeted contents from verified accounts: while most of the retweeted accounts by bots belong to MEDIA group (e.g. \emph{La Repubblica}, \emph{Agenzia Ansa}, \emph{Sky TG 24}), the most retweeted people are \emph{Roberto Burioni} and \emph{Giorgia Meloni}. In general, bots interact most with the MEDIA group; then, they followed the verified accounts of M5S, DX, IV, PD and FI.

\begin{figure}[h!]
    \centering
   \includegraphics[scale=0.7]{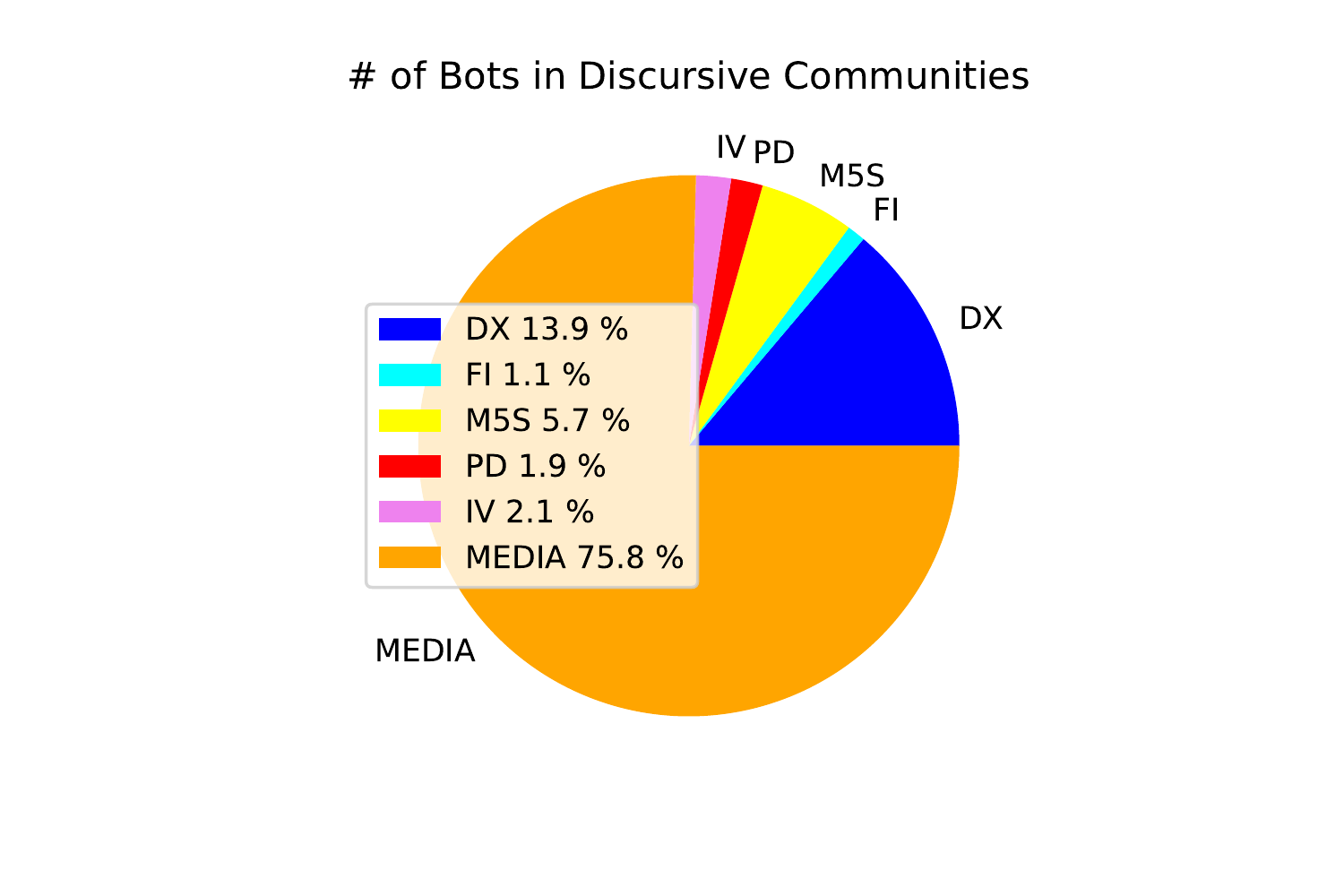}
    \caption{\textbf{Distribution of social-bots in the various discursive communities} As mentioned in the main text, the main contribution to the number of automated accounts comes from MEDIA community, followed by DX and M5S communities.}
    \label{Figure11}
\end{figure}

The community with the largest percentage of social-bots is the FI community with 21.9\%, followed by IV with 10\%, PD with 8.4\%, M5S with 6.5\%, MEDIA with 5.3\% and DX with 4.1\%. When absolute numbers are considered, a strong prevalence of bots from MEDIA community appears, followed by those of the DX community, see Fig.~\ref{Figure11}. More details about the temporal evolution of the activity of automated accounts can be found in the Appendix~\ref{app:temp_bot}.

\subsection{The semantic network}\label{ssec:SemNet}

Let us now analyse the evolution of the narratives characterizing the online debate during the first peak of the contagion. As in~\cite{Radicioni2020,radicioni2021networked}, we start from the bipartite network of accounts and hashtags, where a link between the user $u$ and the hashtag $h$ is present if user $u$ has used hashtag $h$ at least once. Then, by using the same procedure implemented to determine the discursive communities of the validated accounts, we can extract the (validated) semantic network. As already mentioned in the Introduction, the approach that we follow in the present manuscript is slightly different from the one used in~\cite{Radicioni2020,radicioni2021networked}: there, the authors analysed the semantic network defined by each discursive community, while here we consider the ``global'' semantic network and how the various discursive communities interact with it. The resulting network is formed by 5,666 different hashtags, linked by 90,560 connections.

Interestingly, even if the main topic is not strictly political, the most connected hashtags, i.e. those with the highest values of the degree, refer to political parties and politicians: \#\emph{pd}, \#\emph{oms}, \#\emph{m5s}, \#\emph{lamorgese}, \#\emph{regione}, \#\emph{lazio}, \#\emph{dimaio}, \#\emph{governo}, \#\emph{zingaretti}, \#\emph{mes} and \#\emph{conte}\footnote{Respectively, `Democratic Party', `World Health Organization' (OMS is the Italian acronym), `Movimento 5 stelle' (the most represented party in the Italian Parliament), `Luciana Lamorgese' (Italian interior minister), `region', `Lazio' (the Italian region including Rome, administrated by the Italian Democratic Party), `Luigi Di Maio' (at the time of data collection, Italian foreign minister and leader of the M5S), `government', `Nicola Zingaretti' (Democratic Party secretary at the time of data collection and present governor of Lazio), `European Stability Mechanism' (the above mentioned ESM) and `Giuseppe Conte', Italian president of the council of ministers at the time of data collection.}.

%
%

\begin{figure}
    \centering
    \includegraphics[scale=0.6]{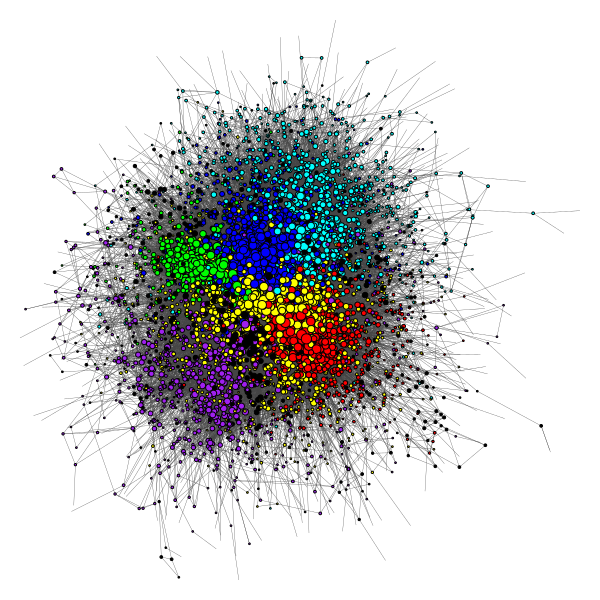}
    \caption{
     \textbf{The validated semantic network: each node represents a different hashtag}. We assigned a different color to the six most numerous communities, which are described in the main text. The dimension of the nodes is proportional to their degree. The network is represented using Fruchterman-Reingold layout.}
    \label{Figure8}
\end{figure}

We, then, run the Louvain algorithm again to detect the various semantic communities; the algorithm provided 61 different communities of hashtags (with modularity $Q\simeq0.56$). We just focused on the most populated ones (see Fig.~\ref{Figure8}). The biggest communities refers to some of the most debated themes and subjects during the pandemic, in particular:\\

\begin{itemize}
    \item the Red community contains mostly political subjects: here we can find the name of the governing political parties at the moment of the data collection. In this sense, heavy criticisms towards the Prime Minister are present;
    \item the White community includes subjects related to the Catholic Church and to the Pope Francis I;
    \item the Yellow community is the most crowded one with pieces of news related to either the local (e.g. at regional level) or the global response to the epidemic;
    \item in Blue, we find updates of the Covid-19 situation (number of deaths, number of contagions, etc.);
    \item the Cyan community includes hashtags related to  trade unions, remote working and to actions adopted by the government at the time of data collections to sustain the employment. Those arguments quickly became hot topics, since the Covid-19 had a heavy impact on the employment and forced firms to take countermeasures such as remote workings~\cite{Patuelli2021};
    \item the Green community includes hashtags related to sports - in particular football, the most followed sport in Italy - that, as many other activities, had to stop.\\
\end{itemize}

More details about the hashtags in the various communities of the semantic network can be found in Appendix~\ref{app:comm}.

\subsubsection{Temporal activities over the semantic network}

\begin{figure}[t!]
    \centering
    \includegraphics[scale=0.6]{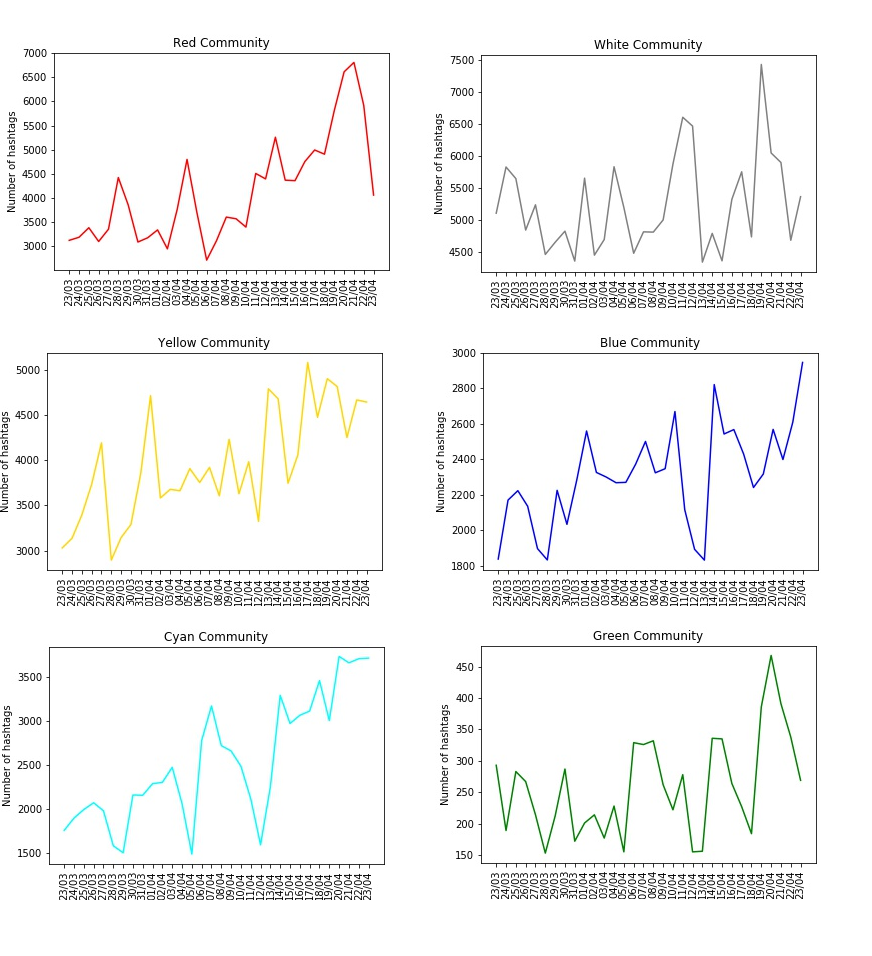}
    \caption{\textbf{Temporal evolution of the number of hashtags shared day to day and belonging to each community in the semantic network}}
    \label{Figure9}
\end{figure}

After the identification of the most important topics within the main communities, we examined the temporal evolution, on a daily scale, of the number of published hashtags belonging to each community (see Fig.~\ref{Figure9}). Tracking these temporal behaviours is important for understanding which events may have caused an increasing Twitter activity about a specific topic. By looking to the peaks in the temporal evolution, the first thing that catches the eye is that all the trends are upwards, indicating that the Twitter conversations about the Covid-19 became more intense since middle April. We identified some events in specific days, which are strictly related to the main topic of the community in exam:\\

\begin{itemize}
    \item\textbf{17/04/2020}: in this day there was the vote on the activation of ``corona bonds'' (joint debt issued to member states of the EU) at the European Parliament. The parties of \emph{Lega Nord} and \emph{Forza Italia} voted against and this caused a lot of comments also on Twitter. There is a peak in this day in the yellow community, which contains also the hashtags \#\emph{coronabond}, \#\emph{eurobond}, \#\emph{lega} and \#\emph{salvinisciacallo}\footnote{Literally, `Salvini jackal', a clear accusations against the leader of Lega Nord, Matteo Salvini, to take advantage of the critical situation to increase his consensus by causing unnecessary difficulties to the government against the interest of the nation.}.
    \item\textbf{19/04/2020}: in this day there is a peak in the curve of the white community. In that day the hashtag \#\emph{25aprile} (i.e. the Italian liberation day from nazi-fascist occupation) was published many times. In particular, some statements of the senator \emph{Ignazio La Russa}, about the nature of the commemorations in that day, caused debates and controversies on Twitter\footnote{The Senator Ignazio La Russa, Fratelli d'Italia, proposed to convert 25th of April from the Liberation (from nazi-fascism) day to the ``day of the commemoration of all the victims of all wars, included the one against the Coronavirus". Due to his membership to a right-wing party, this proposal was read as a tentative to minimize the role of nazi-fascist regime in Italy.}. Indeed, also the hashtag \#\emph{ignaziolarussa} is contained in the white community. 
    \item\textbf{20/04/2020}: the so-called ``second phase'', during which less severe measures have been implemented and some shops and workplaces reopened, started this day. The cyan community contains the hashtags \#\emph{fase2} (phase 2) and consequentially its trend shows a peak on 20/04. In this day a tweet of the US President at the time of data collection, i.e. Donald Trump, declared that he would sign an executive order for suspending immigration in the United States to stop the virus and this announcement caused comments and debates on social networks. Again in the yellow community, which contains also the hashtag \#\emph{trump}, there is a peak in this day. Also the green community shows a maximum on the 20th of April, due to the statements of Sport Minister Vincenzo Spadafora in which he expressed doubts about the resumption of the Italian football championship Serie A.
    \item \textbf{21/04/2020}: the temporal evolution of the red community, i.e. the ``political'' one, shows a peak in this day. The most-shared hashtag of this community, in this day, is \#\emph{quartarepublica}, an Italian TV-program. The day before, the main host of the program was Silvio Berlusconi, leader of the ``Forza Italia'' party and former Prime Minister. Other hosts were the Governor of Veneto region Luca Zaia, the Councilor for Welfare of the Lombardy Giulio Gallera and the Mayor of Naples Luigi De Magistris. As observed in similar studies~\cite{Radicioni2020,radicioni2021networked}, right and center-right wing users are particularly active on mediated events, i.e. public events as TV interview that are heavily covered by users on Twitter.
\end{itemize}

\subsubsection{Semantic activity of the discursive communities}

\begin{figure}
    \centering
    \includegraphics[scale=0.7]{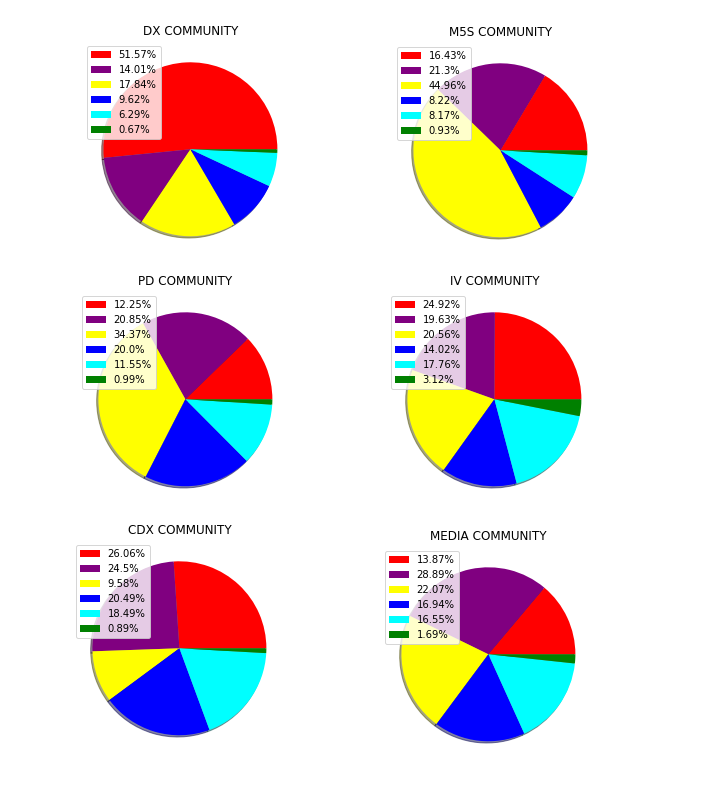}
    \caption{\textbf{Hashtag consumption by discursive community} The pie-charts describe the numbers of times a hashtag, belonging to one of the 6 communities described in the main text, has been shared by a user (both verified and non-verified ones) of the 6 discursive communities identified in the first section.}
    \label{Figure_10b}
\end{figure}

Fig.~\ref{Figure_10b} shows the semantic activity of the various discursive communities, including both verified and unverified users. In the DX group (and, similarly, also in the FI group) there is a sharp prevalence of the red community, due to the presence of hashtags against the government. For what concerns the M5S and the PD, the yellow community is the most shared one. Within the MEDIA group, hashtags are homogeneously distributed among white, yellow, blue and cyan communities.

We repeated the same analysis for social-bots and observed that hashtags by bots are quite evenly distributed across all communities (with the exception of the green one): still, social-bots interact mostly with the verified accounts of MEDIA group and, therefore, they share news and contents of different types. In general the situation changes when considering only verified accounts: more details can be found in Appendix~\ref{app:dynam_ver}. 

\subsubsection{Tracking conspiracy theories and d/misinformation campaigns}

In the literature, ``disinformation" and ``misinformation" have different meanings, both referring to the spread of false information: the former concerns deliberate diffusion, while the latter refers to an unintentional mechanism~\cite{Bradshaw2017a,Bradshaw2018b}. At the present level, we cannot distinguish between the two different natures, thus we always use the term d/misinformation. One of the most interesting aspects of our analysis is the possibility of investigating the spread of forms of d/misinformation online. We have identified 2 sub-communities of hashtags related to d/misinformation campaigns regarding the origin and the diffusion of the coronavirus.

Looking at the hashtags of the first community, connections between Bill Gates, vaccines, 5G, nano-/micro-chips and naturally the Coronavirus, emerge\footnote{This community includes the following hashtags: \#\emph{5G}, \#\emph{5Gkills}, \#\emph{antiinfluenzale}, \#\emph{appimmuni}, \#\emph{bendingspoons}, \#\emph{bigpharma}, \#\emph{bill}, \#\emph{billgates}, \#\emph{billgatesfoundation}, \#\emph{colao}, \#\emph{deepstate}, \#\emph{immuni}, \#\emph{microchip}, \#\emph{montagneir}, \#\emph{nanochip}, \#\emph{no5G}, \#\emph{noimmuni}, \#\emph{spygate}, \#\emph{stop5G}, \#\emph{telecontrollo}, \#\emph{vaccino}, \#\emph{vittoriocolao}.}. Indeed, one of the most widespread false claims seems to be the theory for which the pandemic is a plan masterminded by Bill Gates to implant microchips into humans along with a Coronavirus vaccine. Other interesting connections are those between the hashtags \#\emph{colao} and \#\emph{montagneir}: the first refers to Vittorio Colao, ex CEO of Vodafone and new director of the task force formed by the premier Giuseppe Conte, and the latter refers to the Nobel prize for medicine in 2008, Dr. Montagneir. Conspiracy theorists extracted one phrase from a 2019 video, in which Colao said something about a ``medical substance'' that could be injected thanks to 5G. Instead, Dr. Montagneir stated in an interview that the spread of Coronavirus was a human error by scientists trying to develop a vaccine, precisely against AIDS (in other sub-communities we found also hashtags like \#\emph{HIV} and \#\emph{AIDS}). There are also some references to ``Immuni'' App, which is the application developed by Bending Spoons company, appointed by the Italian government for contact-tracing in order to control the spread of the epidemic; shortly after the release of the app, there were worries about privacy and some users argued that the app was created for spying people.

\begin{figure}[t!]
\centering
\includegraphics[scale=0.67]{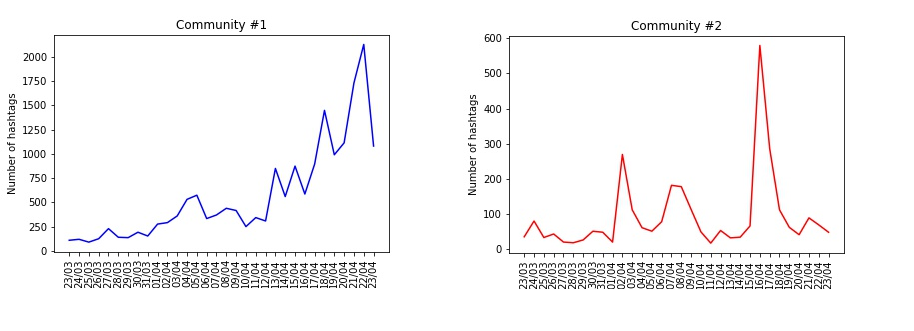}
\caption{\textbf{Temporal evolution of the number of hashtags published day to day and belonging to the two d/misinformation communities described in the text} The various peaks can be related to various offline events that foster the discussion on the various arguments. In the ``5G" hashtag community (left panel) the peak in the end of April is related to the trial in the Hague against the Dutch government for by the group ``Stop5GNL" against the Dutch government. In the ``China" community (right-panel) the peak is related to the publication of a news report about the US Intelligence investigation on the laboratory origin of the Coronavirus.}
\label{Figure12}
\end{figure}

The second community is about the creation in laboratory of the virus by Chinese scientists\footnote{This community includes the following hashtags: \#\emph{china}, \#\emph{chinacentric}, \#\emph{chinaliedpeopledied}, \#\emph{chinamustpay}, \#\emph{chinavirus},\#\emph{chinesecoronavirus}, \#\emph{coronaviruscure}, \#\emph{harvarduniversity}, \#\emph{trumpderangementsyndrome}, \#\emph{who}.}.
We plotted the temporal evolutions of the daily number of published hashtags, also for these communities in Fig. \ref{Figure12}, trying again to identify those events that may have caused an increasing attention on Twitter about these topics. For the community about 5G there is a peak on April 21st and 22nd; on the former there was a trial in the Hague court against the Dutch government for the introduction of 5G brought by the group ``Stop 5GNL''. Moreover, rumors about Bill Gates started to spread from middle April, when conspiracy theorists used a ``TED Talk'' from 2015 in which Gates warned that the world is not prepared for an epidemic, to confirm their theories. The second community has a peak on the 16th of April when a news report was published about the American intelligence investigating the alleged creation of the Coronavirus in laboratory in Wuhan.

Confirming the results of Ref.~\cite{Celestini2020}, those conspiracy communities represent a minority in the semantic network: the hashtags included in the conspiracy communities are respectively 62 and 15, over a total of 5,666 different hashtags. Even in terms of their popularity, their impact is limited: the first conspiracy community was shared 10,452 times and the second one 1,287, against a total of nearly 602,299 messages containing at least one hashtag.

\begin{figure}[t!]
\centering
\includegraphics[scale=0.7]{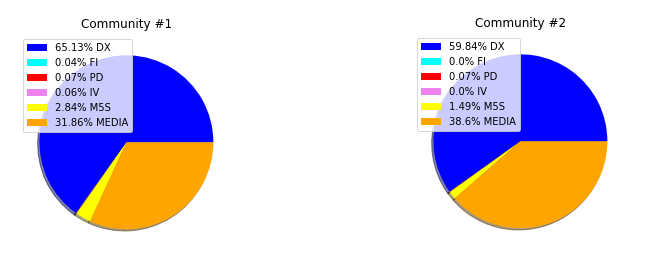}
\caption{
     \textbf{Discursive communities and semantic conspiracy communities} In the plots the number of times users of the discursive communities have published a hashtag belonging to the two different communities of d/misinformation is reported: DX community is the one mostly exposed to conspiracy hashtags.}
\label{Figure13}
\end{figure}

Remarkably, not all discursive communities share conspiracy hashtags in the same way. In Fig.~\ref{Figure13} there are the fractions of users of the 6 different discursive communities listed in the previous paragraph, which shared the hashtags of the three communities of d/misinformation. The DX community is the one most affected by d/misinformation, followed by MEDIA group which however contains much more users than the DX one. Even in this sense, our analysis confirms the findings of Ref.~\cite{Caldarelli2020b}: there the Non Reliable sources, as tagged by \href{https://www.newsguardtech.com/}{NewsGuard}, were almost exclusively shared by DX community (in Ref.~\cite{Caldarelli2020b} MEDIA community was not analysed). 

\section{Methods}\label{sec:Meth}

\subsection{Data}

The data set analysed is similar to the one used in~\cite{Caldarelli2020b}. Data were collected using Twitter Search API across a period of 32 days, between the 23th of March 2020 to the 23th of April 2020, that was one of the most crucial period of the quarantine. In particular we looked at those tweets containing at least one of the following keywords or hashtags: \emph{coronavirus}, \emph{coronaviruses}, \emph{ncov}, \emph{ncov2020}, \emph{ncov2019}, \emph{covid2019}, \emph{covid-19}, \emph{SARS-CoV2}, \emph{\#coronavirus}, \emph{\#WuhanCoronavirus}, \emph{\#coronaviruschina}, \emph{\#CoronavirusOutbreak}, \emph{\#coronaviruswuhan}, \emph{\#ChinaCoronaVirus}, \emph{\#nCoV}, \emph{\#coronaviruses}, \emph{\#ChinaWuHan}, \emph{\#nCoV2020}, \emph{\#nCoV2019}, \emph{\#covid2019}, \emph{\#covid-19}, \emph{\#SARS\_CoV\_2}, \emph{\#SARSCoV2}, \emph{\#COVID19}.

The total number of tweets selected in this way was approximately 1.5 millions, posted by 126,614 different users. Then, we also took all the hashtags in each tweet for identifying the central topic. As a consequence, only tweets containing, at least, one hashtag have been retained, resulting in 602,299 messages. In presence of retweets, we looked also for the user IDs and for the hashtags of the original tweets.

At this point, we had to deal with the problem of orthographic errors or, in general, with all the possible variations of a word that have to be considered as an unique one. For this reason we had to clean our data set, identifying, in some way, those hashtags that were ``similar enough''. We used the Levenshtein distance, that is a sequence-based similarity: it quantifies the cost of transforming a string $x$ into a string $y$ when the two strings are viewed as sequences of characters. In particular we counted the number of characters that had to be changed and we divided it for the total number of characters of the words in exam. We have set a threshold at 0.82 to consider any two strings as the same and kept the one with larger occurrence in the data set. We chose this threshold after several sample checks; this value seemed to be the most effective to identifying the different versions of the same word, reducing the number of different hashtags by about 30\%. Afterwards, we put the information about the ID of the users and their hashtags into the two layers of a bipartite network; a link between a hashtag and an user exists when that user has published a tweet containing that hashtag at least once.

\subsection{Entropy-based null-models as benchmarks}

Once the information about the hashtag usage of the system is represented via a bipartite network, we need to extract signals that cannot be explained only invoking a random usage of hashtag by users. In this sense, we need a proper benchmark: entropy-based null-models are, by construction, unbiased; so, they represent a natural choice~\cite{Cimini2018}. In the present case, we want to extract the various narratives, as identified by group of hashtags, that cannot be explained only 1) by the attitude of users to use different hashtags and 2) by the virality of the latter ones. Since this information is encoded into the degree sequence of both layers, we need a benchmark discounting the information carried by the degree sequence. The entropy-based null-model discounting it is the Bipartite Configuration Model,~\cite{Saracco2015a}). In the present section we will briefly revise the steps to define this null model-

Let us consider a bipartite network in which the two layers $\top$ and $\bot$ have dimension $N_\top$ and $N_\bot$; in the following, Latin indices will be used to identify nodes on the $\top$ layer while Greek ones will be used for the $\bot$ layer. Then, the bipartite network can be represented by its biadjacency matrix, i.e. a $N_\top\times N_\bot$ matrix $\mathbf{M}$ whose generic entry $m_{i\alpha}$ is $1$ if the node $i\in\top$ is connected to the node $\alpha\in\bot$.

First, let us define a \emph{statistical ensemble} of graphs, i.e. the set of all the possible bipartite graphs having the same number of nodes but with different topology, from the fully connected to the empty ones. Then, we can define the Shannon entropy over the ensemble, by assigning a different probability to each element of it:

\begin{equation*}
        S=-\sum_{G_\text{Bi}\in\mathcal {G}_\text{Bi}}P(G_\text{Bi})\ln{P(G_\text{Bi})};
\end{equation*}
here, $P(G_\text{Bi})$ is the probability of the bipartite graph $G_\text{Bi}$. Let us now maximise the entropy, while constraining the network degrees: in particular, we want that the ensemble average of degrees to match their observed value, in order to have a null-model tailor on the real system. In term of the biadjacency matrix, the degree sequences of the $\top$ and $\bot$ layers respectively read $k_i=\sum_\alpha m_{i\alpha}$ and $k_\alpha=\sum_i m_{i\alpha}$. Using the method of the Lagrangian multipliers, constrained maximisation can be expressed as the maximisation of $S'$, defined as

\begin{eqnarray*}
    S'&=&S\nonumber\\
    &&+\sum_i\eta_i\left[k_i^*-\sum_{G_\text{Bi}\in\mathcal {G}_\text{Bi}}P(G_\text{Bi})k_i(G_\text{Bi})\right]+\sum_\alpha\theta_\alpha\left[h_\alpha^*-\sum_{G_\text{Bi}\in\mathcal {G}_\text{Bi}}P(G_\text{Bi})h_\alpha(G_\text{Bi})\right]\nonumber\\
    &&+\zeta\left[\sum_{G_\text{Bi}\in\mathcal {G}_\text{Bi}}P(G_\text{Bi})-1\right]
\end{eqnarray*}
where $S$ is the Shannon entropy defined above, $\eta_i$, $\theta_\alpha$ are the Lagrangian multipliers relative to the degree sequences and $\zeta$ is the one relative to the probability normalization; quantities marked with an asterisk $*$ indicate quantities measured on the real network.

Maximising $S'$ leads to a probability per graph $G_\text{Bi}\in\mathcal{G}_\text{Bi}$ that can be factorised in terms of the probabilities per link $p_{i\alpha}$~\cite{park2004statistical}, i.e.

\begin{equation}
\label{factorized}
    P(G_\text{Bi})=\prod_{i,\alpha}p_{i\alpha}^{m_{i\alpha}(G_\text{Bi})}\,(1-p_{i\alpha})^{1-m_{i\alpha}(G_\text{Bi})},
\end{equation}
where $p_{i\alpha}=\dfrac{e^{-\eta_i-\theta_\alpha}}{1+e^{-\eta_i-\theta_\alpha}}$. Nevertheless, at this level the above equation is just formal, since we do not know the numerical value of the Lagrangian multipliers. To this aim, we can then maximise the likelihood of the of the real network~\cite{Garlaschelli2008,squartini2011analytical}; it can be shown that the likelihood maximisation is equivalent at imposing

\begin{equation*}
    \langle k_i\rangle=k_i^*,\,\forall\:i\in\top;\qquad\langle h_\alpha\rangle=h_\alpha^*,\,\forall \alpha\:\in\bot.
\end{equation*}

\subsection{Validated projection of bipartite networks}

Once we have a well-grounded benchmark, we want to infer similarities among nodes on the same layer. We can use as a measure of similarity the number of common neighbours - for each couple of hashtags, the number of users that have shared both. Let us assume, without loss of generality, that we want to project the information contained in the bipartite network onto the $\top$ layer and call $V_{ij}$ the number of common neighbors between nodes $i,j\in\top$\footnote{Following Ref.~\cite{Saracco2016}, we use the letter $V$ to indicate common neighbours, since this pattern appear in the bipartite network as a \emph{V} between the layer.}

%
%

In terms of the biadjacency matrix, $V_{ij}$ can be expressed as

\begin{equation*}
V_{ij}=\sum_\alpha V_{ij}^\alpha=\sum_\alpha m_{i\alpha}m_{j\alpha},
\end{equation*}
where we have defined $V_{ij}^\alpha= m_{i\alpha}m_{j\alpha}= 1$, if both $i$ and $j$ are connected to node $\alpha\in\bot$. Let us now compare the observed numbers of co-occurrences between each possible pair of nodes in $\top$ with the prediction of the BiCM. Since link probabilities are independent, the presence of each V-motif $V_{ij}^\alpha$ can be regarded as the outcome of a Bernoulli trial:

\begin{equation*}
\begin{split}
f_{\text{Ber}}(V_{ij}^\alpha=1)=&p_{i\alpha}p_{j\alpha},\\
f_{\text{Ber}}(V_{ij}^\alpha=0)=&1-p_{i\alpha}p_{j\alpha}.
\end{split}
\end{equation*}

In general, the probability of observing $V_{ij}=n$ can be expressed as a sum of contributions, running on the n-tuples of considered nodes (in this case, the ones belonging to the layer of users). Indicating with $A_n$ all possible nodes n-tuples among the layer of $\bot$, this probability amounts at

\begin{equation}\label{eq:PB}
f_{PB}(V_{ij}=n)=\sum_{A_n}\left[\prod_{\alpha\in A_n}p_{i\alpha}p_{j\alpha}\prod_{\alpha'\notin A_n}(1-p_{i\alpha'}p_{j\alpha'})\right],
\end{equation}
where the second product runs over the complement set of $A_n$, that is the set of users not belonging to the tuple chosen. Eq.~(\ref{eq:PB}) represent the generalization of the usual Binomial distribution when the single Bernoulli trials have different probabilities, also known as Poisson Binomial distribution~\cite{Hong2013}.

We can, then, verify the statistical significance of the observed co-occurrences by calculating their p-value according to the distribution in~(\ref{eq:PB}), i.e. the probability of observing a number of co-occurrences greater than, or equal to, the observed one:

\begin{equation}
\text{p-value}\big(V^*_{ij}\big)=\sum_{V_{ij}\ge V^*_{ij}}f_{PB}\big(V^*_{ij}\big).
\end{equation}

Repeating this calculation for every pair of nodes, we obtain $\binom{N_\top}{2}$ p-values. In order to state the statistical significance of the hypotheses belonging to this group, it is necessary to adopt a multiple hypothesis testing correction; in the present paper, we use the \emph{False Discovery Rate} (FDR,~\cite{benjamini1995controlling}), since it allows the tolerable percentage of false positives to be tuned.

\subsection{Modularity and community detection}

In the present analysis, we inferred the discursive communities from the communities in the validated network of verified users. In particular, we used the modularity based Louvain algorithm~\cite{Blondel2008}.\\

The modularity~\cite{Newman2010} compares the number of edges within the actual communities with the number of edges one would expect to have in the same communities but regardless of communities structure; the latter quantity depends critically on the chosen null-model. Modularity can be written as

\begin{equation}
    Q=\frac{1}{2m}\sum_{ij}\big(A_{ij}-P_{ij}\big)\,\delta(C_i,C_j)
\end{equation}
where $m$ is the total number of links of the network, $A_{ij}$ are the entries of the adjacency matrix, $P_{ij}$ is the probability to have a link between nodes $i$ and $j$ according to the chosen null-model and the term $\delta(C_i,C_j)$ selects all the pairs of nodes contained in the same community (equal to 1 if $C_i=C_j$ or 0 otherwise). In the original definition in~\cite{Girvan2002} the null model chosen is the Chung-Lu one~\cite{Chung2002}, which conserve the degree sequence, but it is known to be inconsistent for dense networks that present strong hubs. In the present paper we use instead the entropy-based Undirected Configuration Model (UCM) defined in~\cite{Garlaschelli2008,squartini2011analytical}: it can be shown that in the case of sparse network, the UCM can be approximated by the Chung-Lu null-model. In the present case, we implemented the Louvain algorithm using the exact null-model via the Python module \href{https://pypi.org/project/NEMtropy/}{\texttt{NEMtropy}}, described in~\cite{Vallarano2021}.\\

As a last observation, we would like to stress that our dataset is the one considered in~\cite{Caldarelli2020b} and the methodology employed to analyse it is the same. Nevertheless, here we implement the full BiCM, by using the Python module \href{https://pypi.org/project/NEMtropy/}{\texttt{NEMtropy}} described in~\cite{Vallarano2021}; in~\cite{Caldarelli2020b}, instead, the sparse network approximation was used. This may let small disagreements in the composition of the communities found in the present manuscript and of those found in~\cite{Caldarelli2020b} to emerge. In particular, the right-wing community in~\cite{Caldarelli2020b} had more elements and also included the supporters of the \emph{Forza Italia} party - now separated from those of \emph{Lega Nord} and \emph{Fratelli d'Italia}.

\section{Discussion}\label{sec:Disc}

In the present manuscript, we analysed the narratives emerged on Twitter during the Covid-19 epidemic in Italy. Our data set covered a period of one month (from the 23rd of March 2020 to the 23rd of April 2020) and included approximately 1.5 millions of tweets. As observed in previous studies~\cite{Caldarelli2020b}, even in the (non strictly) political discussion concerning the Covid-19, the discursive communities reflects political groups. This behaviour is probably due to the pre-existence of discursive communities: if the accounts followed by a specific users have a certain political orientations, this initial bias will influence even other -even non necessarily political- discussions, just because Twitter platform will display on the user home the updates coming from her/his friends.\\
In the present paper, we have refined the partition in communities obtained in~\cite{Caldarelli2020b} by using the full BiCM, implemented into the Python module \href{https://pypi.org/project/NEMtropy/}{\texttt{NEMtropy}} and described in~\cite{Vallarano2021}, instead of its approximation; not surprisingly, the recovered partition is reflected into the organization of the online debate.

We represent it as a bipartite network of users and hashtags, where a link connects a user to an hashtags if the former one has used the latter one at least once during the observation period. Then, using a bipartite entropy-based null-model~\cite{Saracco2015a} as a benchmark, we have projected the original bipartite network onto the layer of hashtags~\cite{Saracco2016a}. Our approach is similar to the one used in~\cite{Radicioni2020,radicioni2021networked} but with a crucial difference: there, the various communication strategies developed by the different groups during political events were studied by obtaining a semantic network per discursive community; here, we have limited ourselves to consider just one semantic network for the entire debate.

As observed in~\cite{Radicioni2020,radicioni2021networked}, the various discursive communities have a different behaviour: in Fig.~\ref{Figure_10b} we showed that, while the discursive communities closer to the government at the time of the data collection (i.e. M5S, PD and IV) focus more on the news related to the pandemic, the ones closer to the opposition (i.e. DX and FI) focus more on political themes, with the aim of highlighting the inefficiency of the measures adopted by the government to contrast the pandemic. Other behaviours of the various discursive communities observed in~\cite{Radicioni2020,radicioni2021networked} are confirmed by the present study. For instance, \emph{mediated} events stimulate the discussions on Twitter more in DX and FI community: by looking at the temporal evolution of the communities of the semantic network, we were able to identify these events, as TV-appearances of politicians.

Confirming other previous studies~\cite{Celestini2020}, the communities of hashtags related to d/misinformation represent a limited number of hashtags. Nevertheless, it is striking that they are mostly ``visited'' by a single discursive community, i.e. the right-wing one, as shown in Fig.~\ref{Figure13}. Indeed, such an impressive percentage of accounts using those hashtags in their messages is not justified by the number of accounts in this discursive community. Such an exposure to d/misinformation campaigns of this discursive community has already been observed in~\cite{Caldarelli2020b}. Remarkably, in this discursive community the incidence of automated accounts is extremely limited, according to Botometer~\cite{Yang:2013}. In a sense, social bots seem to be more focused on ``debating'' subjects of wide interest: they formed the 3\%-4\% of the tweets of the data set and the hashtags they share seem to be distributed quite evenly across the different semantic communities. In fact, they mostly retweeted the contents of MEDIA group; their temporal activity do not suggest particular events which could stimulate their activity on Twitter.

The identification of the various communication behaviours of the Twitter users represents a data-driven tool to measure - and, therefore, elaborate different strategies to limit - the exposure of the users themselves to d/misinformation campaigns. In fact, the presence of groups that are particularly susceptible to low-quality contents limits the efficiency of debunking activities~\cite{Zollo2017}: in this sense, different strategies should be elaborated to effectively fight d/misinformation.


\begin{backmatter}

\section*{Competing interests}

The authors declare no competing interests.

\section{Funding}
GC acknowledges support from ITALY-ISRAEL project Mac2Mic and EU project nr. 952026 - HumanE-AI-Net. FS and TS acknowledges support from the EU project SoBigData-PlusPlus, nr.\ 871042. MM, GC and FS acknowledge support from the IMT School for Advanced Studies PAI project Toffee.

\section{Author's contributions}
MM contributed to the design of the work, to the analysis of the data and to their interpretation. GC, TS and FS contributed to the conception and to the design of the work and to the interpretation of the data. All authors have drafted the work or substantively revised it and have approved the submitted version and have agreed both to be personally accountable for the authors' own contributions and to ensure that questions related to the accuracy or integrity of any part of the work, even ones in which the author was not personally involved, are appropriately investigated, resolved, and the resolution documented in the literature.

 


\bibliographystyle{bmc-mathphys} 











\end{backmatter}

\newpage

\appendix

\section{Description of the political communities}\label{app:polcomm}
The reader not used to the Italian political scenario may need a more detailed description of the cited political parties and of the various characters.\\  

\noindent 
The \emph{Movement 5 Stars} (M5S) is a political party founded on the web in 2009 by the comedian Beppe Grillo as an anti-establishment movement. After getting important results at the local elections (like the election of Virginia Raggi as major of the city of Rome and the election of Chiara Appendino as major of Torino in 2016) the M5S resulted as the most voted party, under the guidance of Luigi Di Maio, at the national elections of 2018. Beside the great result (it was the second time that the M5S was present at a national election), the Movement did not have enough representatives in the parliament to form a government. The political group decided to form a post-election alliance with Lega, the most voted political party of the most voted alliance, i.e the right and center-right one. The government was guided by the prime minister Giuseppe Conte and lasted 1 year and 3 months; in August 2019, the leader of Lega Matteo Salvini to negate the confidence vote to Conte and opened a political crisis. After few weeks of debate, the M5S allied with the Italian Democratic Party (\emph{Partito Democratico}, or PD) to form a novel government, after the openness of the former PD secretary Matteo Renzi. The new government was again guided by Giuseppe Conte; Luigi Di Maio, the leader of the Movement, took the role of Minister of Labour. Another important ministry that was ruled by the Movement was the Ministry of Justice, guided by Alfonso Bonafede. From its foundation, the Movement 5 Stars has been supported by the newspaper \emph{Il Fatto Quotidiano}, guided by Marco Travaglio.\\

\noindent The main \emph{right-wing parties} in the Italian parliament are Lega and Fratelli d'Italia, guided respectively by Matteo Salvini and Giorgia Meloni. Together with Forza Italia, a center-right-wing party guided by former prime minister Silvio Berlusconi), they presented as a right-wing alliance at the national election of 2018. The alliance was the most voted one, but the number of representatives in the parliament was not enough to express a government. Lega, that resulted as the most voted party in the most voted alliances (even if not the most voted party in general), decided to break the previous alliance to make a new one with M5S. The government was guided by Giuseppe Conte; quite obviously, the choice of Matteo Salvini was heavily criticized by the former allies. Lega abandoned the new alliance with M5S in summer 2019 and went to the opposition.\\

\noindent The \emph{Italian Democratic Party} (PD), guided, at the time of the data collection, by Nicola Zingaretti, was the second most voted party at the election at the national elections of 2018. After the alliance with the M5S in summer 2019 to form a new government, PD experienced the split of a group of representatives, guided by the former PD secretary Matteo Renzi, forming the new political party of Italia Viva. Paolo Gentiloni, former prime minister from the end of 2016 to June  2018 and PD parliament representative, was then elected as European Commissioner for Economy. In the first months of 2021, Nicola Zingaretti resigned as PD secretary and was replaced by Enrico Letta.\\

\noindent\emph{Italia Viva} is the political party founded by the former prime minister and former PD secretary Matteo Renzi, soon after the formation of the second government guided by Giuseppe Conte: Italia Viva continued supporting (with two ministers) the government, but marked its difference from PD. Important personalities in Italia Viva are Maria Elena Boschi and Ivan Scalafarotto. As already demonstrated in previous experiences in PD, Matteo Renzi is a capable user of the new technologies as a form of communication.\\

\noindent\emph{Forza Italia} (FI) is the political party founded by the former three-times prime minister Silvio Berlusconi. The performances at the last national elections were particularly unsatisfying for FI, that lost its supremacy in the center-right and right-wing alliance in favour of Lega. Important representative of the party are Renato Brunetta and Antonio Tajani, the national coordinator and former President of the European Parliament.

\section{Polarization of unverified users}\label{app:pol_uu}
In the present appendix we analyse how non-verified users interact with the verified accounts. We followed the procedure shown in Ref.~\cite{Becatti2019}.\\
We denote with $\mathcal{C}_c=1,2,3,4,5,6$ the main 6 different (political) discursive communities of verified users described above and with $\mathcal N_\alpha$ the set of neighbours of a non-verified user $\alpha$ in the bipartite network, i.e. the set of verified users the node $\alpha$ has interacted with. Following~\cite{Bessi2016}, we define the so-called \emph{polarisation index} for the node $\alpha$ as,
\begin{equation}
    \rho_\alpha=\max \big(\{ \mathcal I_{\alpha,c}\,:\,c=1,2,3,4,5,6\}\big)\quad\text{for}\quad\alpha\in U,
\end{equation}
with
\begin{equation}
    \mathcal I_{\alpha,c}=\frac{|\mathcal C_c\cap\mathcal N_\alpha|}{|\mathcal N_\alpha|}\quad\text{for}\quad c=1,2,3,4,5,6,
\end{equation}
and $U$ is the set of unverified users. The term $\mathcal I_{\alpha,c}$ denotes the fraction of interactions of the node $\alpha$ towards community $c$. The index built in this way is bounded in $[0,1]$ and $\rho_\alpha=0$ means that no interaction has been observed with the six groups; for example, $\rho_\alpha$ close to $1/6$ indicates that user $\alpha$ equally interacts with the six communities and $\rho_\alpha=1$ means that it has always interacted with only one group.
\begin{figure}
 \centering
 \includegraphics[scale=0.8]{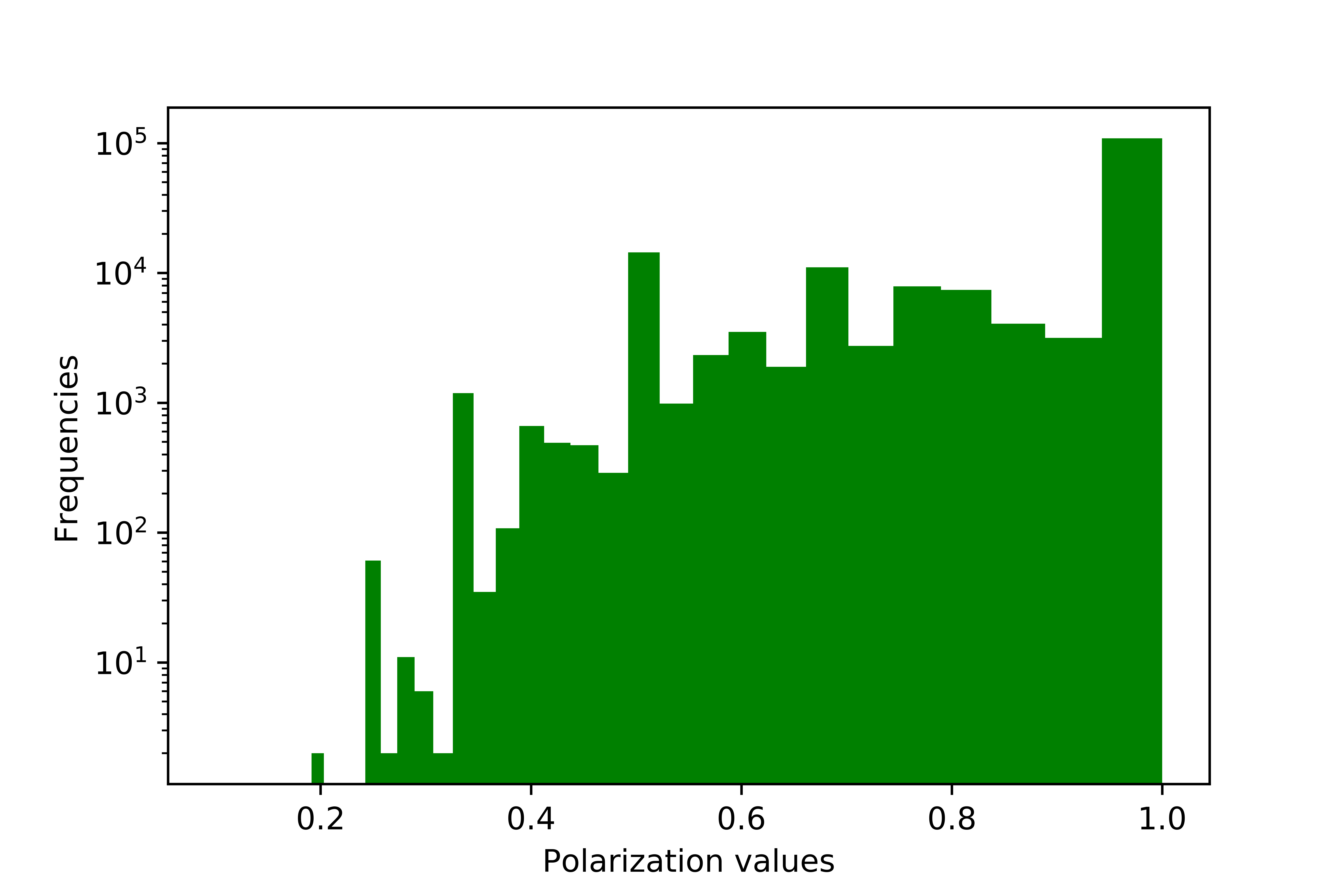}
  \caption{\textbf{Histogram of the polarisation values obtained for non-verified users in logarithmic scale.} Values of polarisation close to 1 indicate that users interact mostly with only one political group. In the graphic the number of users with polarisation between 0.95 and 1 is at least approximately 10 times greater than the frequencies for the other values of polarisation.}
      \label{Figure2}
      \end{figure}
The histogram of the polarisation values obtained for all the non-verified users is displayed in Fig.~\ref{Figure2}. The results indicate that most of the non-verified users has an extremely unbalanced distribution of their interactions with the members of the other alliances, since they mostly retweet contents shared by people from their own community rather than from different ones; indeed, many non-verified users show high polarisation values (near to 1).

\section{Temporal activity of automated accounts per community}\label{app:temp_bot}
\begin{figure}
\centering
\includegraphics[scale=0.8]{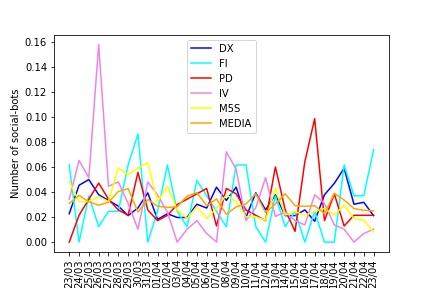}
\caption{\textbf{The temporal evolution of the number of social-bots retweeting the verified accounts of the  different political discursive communities} The values on y-axis are divided, day to day, by the total number of retweets in the entire period for each community.}
\label{Figure4}
\end{figure}

In Fig.~\ref{Figure4} the temporal evolution of the number of times bots retweeted a post by one of the six communities of verified accounts is display. The various trends are normalized by dividing for the total number of retweets for each group in the entire period. 
It is interesting to note the peak on the 26th of March for IV group; in this day bots retweeted many times posts by the virologist Roberto Burioni. Apart from others smaller peaks, the number of bots retweeting verified accounts oscillates around the same values, for all the groups, not suggesting particular events which could have stimulated their activity on Twitter.

\section{Description of the communities in the semantic network}\label{app:comm}

In the present appendix we present a more detailed description of the main communities in the semantic network of Fig.~\ref{Figure8}.

\begin{description}
    \item[Red] 
    The Red community contains 761 different hashtags. The hashtags with the highest degree are names of politicians or parties, like \#\emph{pd}, \#\emph{dimaio}, \#\emph{lamorgese} and \emph{\#zingaretti}\footnote{Democratic Party, Italian foreign Minister Luigi Di Maio, Italian interior Minister Luciana Lamorgese and Democratic party secretary Nicola Zingaretti.}. Using again the Louvain algorithm within this community, we were able to identify specific topics; for instance, in one sub-community we found many hashtags about \emph{immigration} (like \#\emph{migranti}, \#\emph{immigrati}, \#\emph{clandestini}, \#\emph{alankurdi}, \#\emph{ong}\footnote{`Migrants', `immigrants', `Illegal immigrants', the ship `Alan Kurdi' for the rescue of migrants, `Non-Governmental Organization'. }), which has been one of the most discussed political topics in Italy. Especially during the epidemic some extreme right accounts connected the migration from Northern Africa to the spread of the contagions, without reasonable evidences.  There is also a sub-community which contains hashtags of protest against the government, like \#\emph{governocriminale}, \#\emph{governodelcontagio}, \#\emph{contedimettiti}, \#\emph{contebugiardo} and \#\emph{vogliamovotare}\footnote{`Criminal government', `Government of the contagion', `Conte resign', `Conte liar' (the premier Giuseppe Conte), `We want to vote'.}.
    Another popular topic is the European Stability Mechanism (the Italian acronym is MES), which is a European Union agency that provides financial assistance to European countries, in the form of loans or as new capital to banks in difficulty \cite{walker}. Indeed, among the most connected hashtags in this community there are also \#\emph{esm}, \#\emph{mes}, \#\emph{sure}\footnote{The European instrument for temporary ``Support to mitigate Unemployment Risks in an Emergency"  is intended ``to mobilise significant financial means to fight the negative economic and social consequences of the coronavirus outbreak on their territory" (\url{https://ec.europa.eu/info/business-economy-euro/economic-and-fiscal-policy-coordination/financial-assistance-eu/funding-mechanisms-and-facilities/sure_en}).}, \#\emph{nomes} and \#\emph{nosure}.
    \item[White]
    This communities contains 815 elements and the hashtags with the highest degree are all the dates of the period of study (\#\emph{23marzo}, \#\emph{24marzo} and all the others). We identified two main topics in it: there is a group of hashtags referring to the Pope and the Catholic Church in general (\#\emph{papafrancesco}, \#\emph{papa}, \#\emph{messa}, \#\emph{vaticano}, \#\emph{chiesa}\footnote{`Pope Francis', `pope', `Mass', `Vatican', `church'.} and others) and another group of hashtags about TV-programs and films (\#\emph{tvtime}, \#\emph{amici19}, \#\emph{piratideicaraibi}, \#\emph{harrypotter}\footnote{`TV time', the Italian TV-program `amici', the movies `Pirates of the Caribbean' and `Harry Potter'.} and others). 
    \item[Yellow]
    This community has 1159 elements. It is a very heterogeneous community and it was not easy to understand the common topic. Therefore we computed the betweenness centrality for all the nodes of this community and the hashtag \emph{\#regione} (`region') results the most central. Indeed, there are several references to Italian regions like \emph{\#basilicata}, \emph{\#calabrialeghista}, \emph{\#regionesardegna}, \emph{\#regionelombardia}\footnote{`Basilicata', `Calabria governed by Lega Nord', `Sardinia' and `Lombardy'.} and others. In particular most of the hashtags refers to the management of the Coronavirus emergency in Lombardy, the most affected region by the virus in the first wave of the contagion: \#\emph{lombardia}, \#\emph{commissariatelalombardia} (referring to a petition started on Twitter asking the government to put Lombardy under external administration), \#\emph{gallera}, \#\emph{fontana}, \#\emph{fontanadimettiti}, \#\emph{rsa} (in Lombardy the so-called `Residenze Sanitarie Assistenziali' were violently attacked by the virus), \#\emph{milano}\footnote{`Lombardy', the petition described above, assessor for welfare in Lombardy `Giulio Gallera', president of Lombardy `Attilio Fontana', `Fontana resign', `Nursing home', `Milan'.} and others. Another set of vertices seems to refer to foreign countries and cities: \#\emph{francia}, \#\emph{cina}, \#\emph{usa}, \#\emph{giappone}, \#\emph{londra}, \#\emph{regnounito}, \#\emph{newyork}, \#\emph{australia}, \#\emph{russia}, \#\emph{trump}, \#\emph{sanchez}\footnote{`France', `China', `USA', `Japan', `London', `United Kingdom', `New York', `Australia', `Russia', the ex president of the United States Donald `Trump', the president of the Spanish government Pedro `Sanchez'.}  and others.
    \item[Blue]
    This community contains 975 hashtags. It collects those hashtags referring to data and updates about Covid-19. Hashtags with the highest degrees are: \emph{\#casi}, \emph{\#controlli}, \emph{\#decessi}, \emph{\#ricoveri}, \emph{\#calano}, \emph{\#positivi}\footnote{``cases", ``controls``deaths", ``hospitalizations", ``drop", ``positives".} and many others. There are also references to the rules to follow in those days like \emph{\#spostamenti}, \emph{\#chiusi}, \emph{\#comuni}, \emph{\#misure}, \emph{\#negozi}, \emph{\#ordinanza}, \emph{\#sindaco}, \emph{\#sanzioni}\footnote{`moving', `closed', `municipality', `measures', `shops', `order', `mayor', `sanctions'.} and others. Another common topic seems to be economics and therefore all the measures taken and the decrees issued during the pandemic; some of the most connected hashtags are \#\emph{economia}, \#\emph{mercati}, \#\emph{aziende}, \#\emph{credito}, \#\emph{finanziamenti}, \#\emph{imprese}, \#\emph{liquidità}, \#\emph{decretoliquidità}\footnote{`Economics', `markets', `companies', `credit', `financing', `businesses', financial `liquidity' and `liquidity decree'.} and others. 
    \item[Cyan]
    In this community with 966 elements, the main topic seems to be about the employment crisis during the pandemic. Among the hashtags with high degrees there are \#\emph{lavoro}, \#\emph{lavoratori}, \#\emph{sindacati}, \#\emph{ciglcisluil}, \#\emph{pensioni}, \#\emph{imprese}\footnote{`Job, `workers', `labor unions', `pensions', `enterprises'.}. Another topic linked to the previous one regards digital platforms, new technologies and remote working: \#\emph{digitale}, \#\emph{tecnologia}, \#\emph{ai}, \#\emph{innovazione}, \#\emph{online}, \#\emph{robot}, \#\emph{google}, \#\emph{facebook}, \#\emph{web}, \#\emph{smartworking}\footnote{`Digital', `technology', `Artificial Intelligence', 'innovation', `online', `robot', `Google', `Facebook', `Web', `remote working' (during the epidemic many people worked remotely from their homes).}. A special attention was reserved for the topic of online privacy; in particular, with the release of `Immuni' app for contact tracing, a lot of debates emerged about the use of private data by the government. We found a sub-community with hashtags like \#\emph{contacttracing}, \#\emph{cybersecurity}, \#\emph{cybercrime}, \#\emph{dataprotection}, \#\emph{privacy}, \#\emph{security}, \#\emph{garanteprivacy}\footnote{`Guarantor of privacy'.}.
    \item[Green]
    This community contains 318 elements. In this case the main topic is sports, and, in particular, football. In fact, the most connected vertex is the hashtag \#\emph{spadafora} (sports minister) and other hashtags with a high degree are \emph{\#atalanta}, \emph{\#bundesliga}, \emph{\#campionato}, \emph{\#dybala} (he was affected by Coronavirus), \emph{\#fifa}, \emph{\#fiorentina}, \emph{\#galliani}, \emph{\#juventus}, \emph{\#lazio}, \emph{\#liverpool}, \emph{\#seriea}\footnote{`Atalanta F.C.', `Bundesliga', `championship', `Paulo Dybala', `FIFA', `Fiorentina F.C.', `Adriano Galliani', `Juventus F.C.', `Lazio F.C.', `Liverpool F.C.', `Serie A'.} and many others. During the lockdown it has been discussed for a long time the problem of the restart of the European football leagues. There are also references to other sports: \emph{\#basket}, \emph{\#ciclismo}, \emph{\#djokovic}, \emph{\#pallavolo}\footnote{`basketball', `cycling', `Novak Djokovic', `volleyball'.} and many others.\\
\end{description}

\section{Dynamics of verified users over the semantic networks}\label{app:dynam_ver}

\begin{figure}[h!]
    \centering
    \includegraphics[scale=0.7]{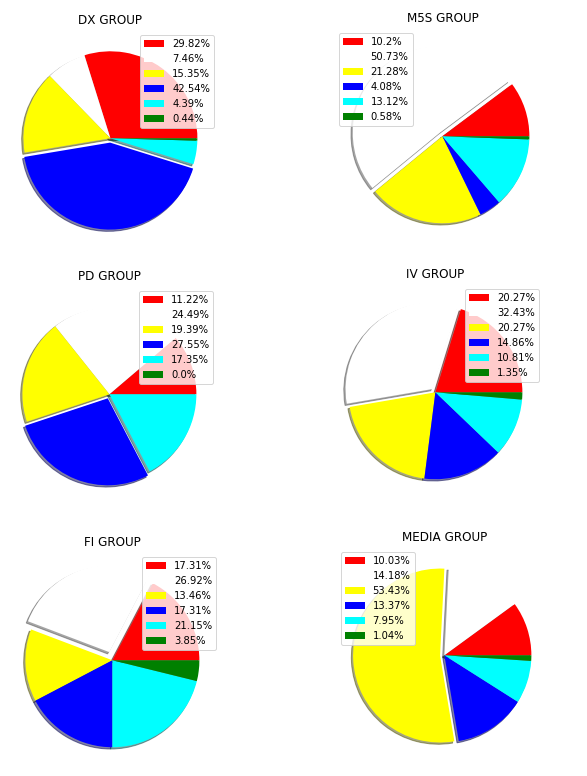}
    \caption{
     \textbf{In the graphics there are the numbers of times a hashtag, belonging to one of the semantic communities described before, has been shared by a verified user of the  discursive communities}}
    \label{Figure10}
\end{figure}

We combined the information on the communities of hashtags in the semantic network with the discursive communities of verified users described in the first section, in order to highlight the activity of verified users in generating new topics. In Fig.~\ref{Figure10} there are the numbers of times a hashtag, belonging to one of the six communities described before, has been shared by a verified user of the 6 discursive communities identified in the Section~\ref{ssec:DiscComm}. The hashtags shared by the smallest groups like PD, IV and FI are more distributed among the communities than those by DX, M5S and MEDIA communities. In detail, DX's members shared mostly hashtag from the blue and the red community, focusing on politics and economics. The verified accounts of M5S shared contents mainly from white community, which is about more general topics, whereas MEDIA group from the yellow one, which refers also to the international scenario and to the situations in the different Italian regions.
\end{document}